\documentclass[sigconf]{acmart}

\usepackage{booktabs} 

\copyrightyear{2022}
\acmYear{2022}
\setcopyright{acmcopyright}\acmConference[WWW '22]{Proceedings of the ACM Web
Conference 2022}{April 25--29, 2022}{Virtual Event, Lyon, France}
\acmBooktitle{Proceedings of the ACM Web Conference 2022 (WWW '22), April
25--29, 2022, Virtual Event, Lyon, France}
\acmPrice{15.00}
\acmDOI{10.1145/3485447.3512117}
\acmISBN{978-1-4503-9096-5/22/04}

\usepackage{balance}
\usepackage{graphicx}
\usepackage{epsfig,epstopdf}
\usepackage{subfigure}
\usepackage{multirow}
\usepackage{booktabs}
\usepackage{color,xcolor}
\usepackage{url}
\usepackage{bm}
\usepackage{enumitem,balance,mathtools}
\usepackage{wrapfig}
\usepackage{euscript}
\usepackage{algorithm}
\usepackage{algorithmic}
\usepackage{ifpdf}
\usepackage{diagbox}
\usepackage{caption}
\usepackage{makecell}
\usepackage{subfigure}
\usepackage[leftcaption]{sidecap}
\usepackage{textcomp}
\usepackage{upgreek}

\usepackage{array}
\newcolumntype{N}{@{}m{0pt}@{}}

\usepackage{lipsum}

\newcommand{\minisection}[1]{\vspace{5pt}\noindent\textbf{#1.}}

\begin{document}
	\title{Learn over Past, Evolve for Future: Search-based Time-aware Recommendation with Sequential Behavior Data}
	\author{Jiarui Jin$^{1,*}$, Xianyu Chen$^{1,*}$, Weinan Zhang$^{1,\dag}$, Junjie Huang$^{1}$, Ziming Feng$^{2}$, Yong Yu$^{1}$.}
\affiliation{$^1$Shanghai Jiao Tong University, China; $^2$China Merchants Bank Credit Card Center, China}
\email{{jinjiarui97, xianyujun, legend0018, wnzhang, yyu}@sjtu.edu.cn, zimingfzm@cmbchina.com}

    \renewcommand{\shortauthors}{J. Jin, et al.}
	\renewcommand{\shorttitle}{STARec}
	\settopmatter{printacmref=false}


	\begin{abstract}
		The personalized recommendation is an essential part of modern e-commerce, where user's demands are not only conditioned by their profile but also by their recent browsing behaviors as well as periodical purchases made some time ago.
		In this paper, we propose a novel framework named \textbf{S}earch-based \textbf{T}ime-\textbf{A}ware \textbf{Rec}ommendation (\textbf{STARec}), which captures the evolving demands of users over time through a unified search-based time-aware model.
		More concretely, we first design a search-based module to retrieve a user's relevant historical behaviors, which are then mixed up with her recent records to be fed into a time-aware sequential network for capturing her time-sensitive demands.
		Besides retrieving relevant information from her personal history, we also propose to search and retrieve similar user's records as an additional reference.
		All these sequential records are further fused to make the final recommendation.
		Beyond this framework, we also develop a novel label trick that uses the previous labels (i.e., user's feedbacks) as the input to better capture the user's browsing pattern.
		We conduct extensive experiments on three real-world commercial datasets on click-through-rate prediction tasks against state-of-the-art methods.
		Experimental results demonstrate the superiority and efficiency of our proposed framework and techniques.
		Furthermore, results of online experiments on a daily item recommendation platform of Company X show that STARec gains average performance improvement of around 6\% and 1.5\% in its two main item recommendation scenarios on CTR metric respectively.
	\end{abstract}
	\settopmatter{printacmref=false} 
	\maketitle
	
{
\renewcommand{\thefootnote}{\fnsymbol{footnote}}
\footnotetext[1]{Jiarui Jin and Xianyu Chen contributed equally to the work.}
\footnotetext[2]{Weinan Zhang is the corresponding author.}
}

{\fontsize{8pt}{8pt} \selectfont
	\textbf{ACM Reference Format:}\\
	Jiarui Jin, Xianyu Chen, Weinan Zhang, Junjie Huang, Ziming Feng, Yong Yu. 2022. Learn over Past, Evolve for Future: Search-based Time-aware Recommendation with Sequential Behavior Data. In \textit{Proceedings of the ACM Web Conference 2022 (WWW '22), April 25--29, 2022, Lyon, France}. ACM, New York, NY, USA, 10 pages.
	\url{https://doi.org/10.1145/3485447.3512117}}

\vspace{-3mm}

\section{Introduction} 
Due to the rapid growth of user historical behaviors, it becomes an essential problem to build an effective recommendation model to help users to find their desired items from a huge number of candidates.
Classical recommendation methods, including collaborative filtering based models \citep{bai2017neural,koren2008factorization,he2017neural} and factorization machine based models \citep{rendle2010factorization,cheng2016wide,juan2016field}, have mainly focused on modeling user's general interests to find her favorite products; while less exploring the user's demands with an aspect of time. 
As stated in \citep{bai2019ctrec,zhu2017next,wu2017recurrent}, time is definitely an important factor that can significantly influence user's demands and result in periodical user behaviors.
Therefore, a branch of recent attempts are proposed to capture user's sequential patterns through either memory networks \citep{ren2019lifelong}, recurrent neural networks \citep{wu2017recurrent,zhu2017next}, or temporal point processes \citep{du2016recurrent,bai2019ctrec}.
However, most existing approaches can only be applied for user behavior data with length scaling up to hundreds due to the computation and storage limitations in real online system \citep{zhou2019deep,zhou2018deep,pi2020search,qin2020user}. 

To tackle this issue, we consider combining it with recently proposed search-based models \citep{pi2020search,qin2020user}, whose key idea is to first search the effective information from  the user's historical records to capture specific interests of the user in terms of different candidate items, which are then used to make the final prediction.
However, it's non-trivial to do that due to the following challenges:
\begin{itemize}[topsep = 3pt,leftmargin =5pt]
\item \textbf{(C1)} How to incorporate the user's sequential patterns into these search-based models?
Existing search-based methods \citep{pi2020search,qin2020user} 
overlook user's sequential patterns (i.e., the effect of time factor).
As a consequence, 
when a teen has purchased a lipstick, these methods are likely to recommend the same or similar products before 
she gets tired of or runs out of the purchased one.
Hence, it's essential to take the time information into account, as the user's demands are highly time-sensitive.
\item \textbf{(C2)} How to leverage the label information (i.e., user feedbacks) from historical data in the recommendation model?
The principal way to use the user historical feedbacks is to treat this feedbacks as the label to supervise the model.
However, as discussed in \citep{wang2021bag,shi2020masked}, combining the information from both label and feature as the input to train the model can significantly improve its performance.
As directly mixing up all this information will definitely lead to the label leakage issue, then how to smartly enrich the model with the label information needs
to investigate.
\item \textbf{(C3)} How to design a learning algorithm to simultaneously train a search-based model and a prediction model in an end-to-end fashion?
Previous attempts 
either manually design a mixed loss function \citep{pi2020search} or apply a reinforcement learning (RL) \citep{qin2020user} in training.
As the performance of the former one largely relies on the loss design and hyper-parameter tuning, and the latter one usually suffers from the sample inefficiency of the RL algorithm, the training algorithm design also is another significant challenge.
\end{itemize}

In this paper, we propose a novel sequential recommendation framework named \textbf{S}earch-based \textbf{T}ime-\textbf{A}ware \textbf{Rec}ommendation (\textbf{STARec}) which captures user's evolving demands over time through a unified search-based time-aware model.

\begin{figure}[t]
	\centering
	\includegraphics[width=1.0\linewidth]{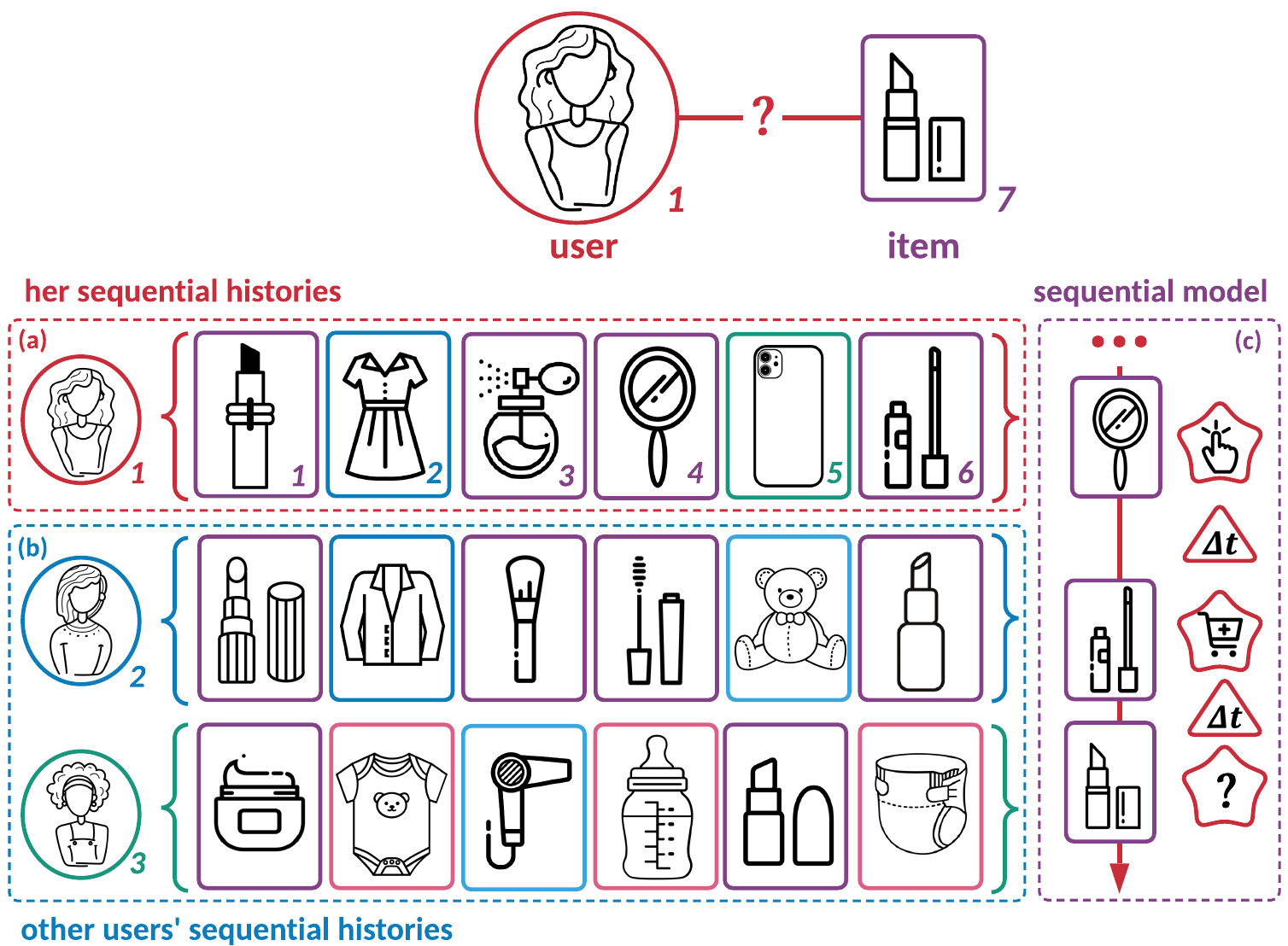}
	\vspace{-7mm}
	\caption{
		An illustrated example for motivations of STARec: For search-based module, (a)
		among historical records of a user $u_1$, 
		we search items (e.g., $i_1$) with the same or similar category to the target item $i_7$; (b) to further enrich the information, we involve similar users' (e.g., $u_2$'s) related histories as reference. For the time-aware module, 
		(c) we develop a 
		sequential network, 
		and design a label trick to involve the user's previous feedbacks as input. In this case, the label of the target item (denoted as $?$) is set as a randomized value.
	}
	\label{fig:motivation}
	\vspace{-4mm}
\end{figure}

Concretely, noticing that 
\emph{category} plays an essential role in search models \citep{pi2020search},
we first construct an embedding vector for each category.
We then search and retrieve items either by a hard-search strategy based on category IDs or a soft-search strategy based on the similarities between their category embeddings.
The intuition of using category for search and retrieval is straightforward. 
Taking Figure~\ref{fig:motivation}(a) as an instance, the motivation of the teen $u_1$ buying the lipstick $i_1$ can either lie in that she is running out of her old lipstick $i_2$, or that she needs an accessory for her new purchases (e.g., lip gloss $i_4$), but not likely due to her purchased iPhone $i_3$.
Note that our search-based module using category embeddings instead of item embeddings would make the whole framework much easier to train.
We also design a novel adaptive search mechanism that can gradually transfer from the hard-search strategy to the soft-search one when the embedding vectors are well-tuned.

In order to mine the hidden time-aware patterns, we then mix up the retrieved items together with recent browsed ones and feed them into a time-aware sequential network that considers not only the sequential orders but also their time intervals.
Besides the user's own histories, we also attempt to enrich the model by similar users' historical records.
As shown in Figure~\ref{fig:motivation}(b), when recommending $i_1$ to $u_1$, we argue that referring to similar historical records from similar users such as $u_2$ would be helpful; while the histories of dissimilar users such as $u_3$ would be noise.
This user similarity can be either softly estimated through the inner-product of their embedding vectors or can be hardly measured by counting the number of purchased items with the same category with $i_1$.

Different from current prevailing methods using user's feedbacks (e.g., click, purchase) only as the supervision signals.
As Figure~\ref{fig:motivation}(c) shows, we propose to involve the user's previous feedbacks as input, where the label of the target item is set as a randomized value.
We call this technique \emph{label trick}.
Its intuition is straightforward that if a user finds her desired items, it's unlikely for her to click or purchase other similar items. 


In summary, the contributions of the paper are three-fold:
\begin{itemize}[topsep = 3pt,leftmargin =5pt]
\item We propose a novel framework named STARec, which captures the user's time-sensitive evolving demands via combining a search-based module and a time-aware module.
\item We propose to involve the user's previous feedbacks as input and reveal that this label information can improve the performance.
\item We design a new adaptive search mechanism, which gradually transfers from the hard-search strategy to the soft one. 
\end{itemize}

We conduct extensive experiments on three industrial datasets, and experimental results exhibit that the superiority of STARec over the state-of-art methods. 
We successfully deploy STARec in two main item recommendation scenarios in Company X, and share our hands-on experience and discuss the potential extensions to ranking tasks in the appendix.

\begin{figure*}[t]
	\centering
	\vspace{-2mm}
	\includegraphics[width=1.00\linewidth]{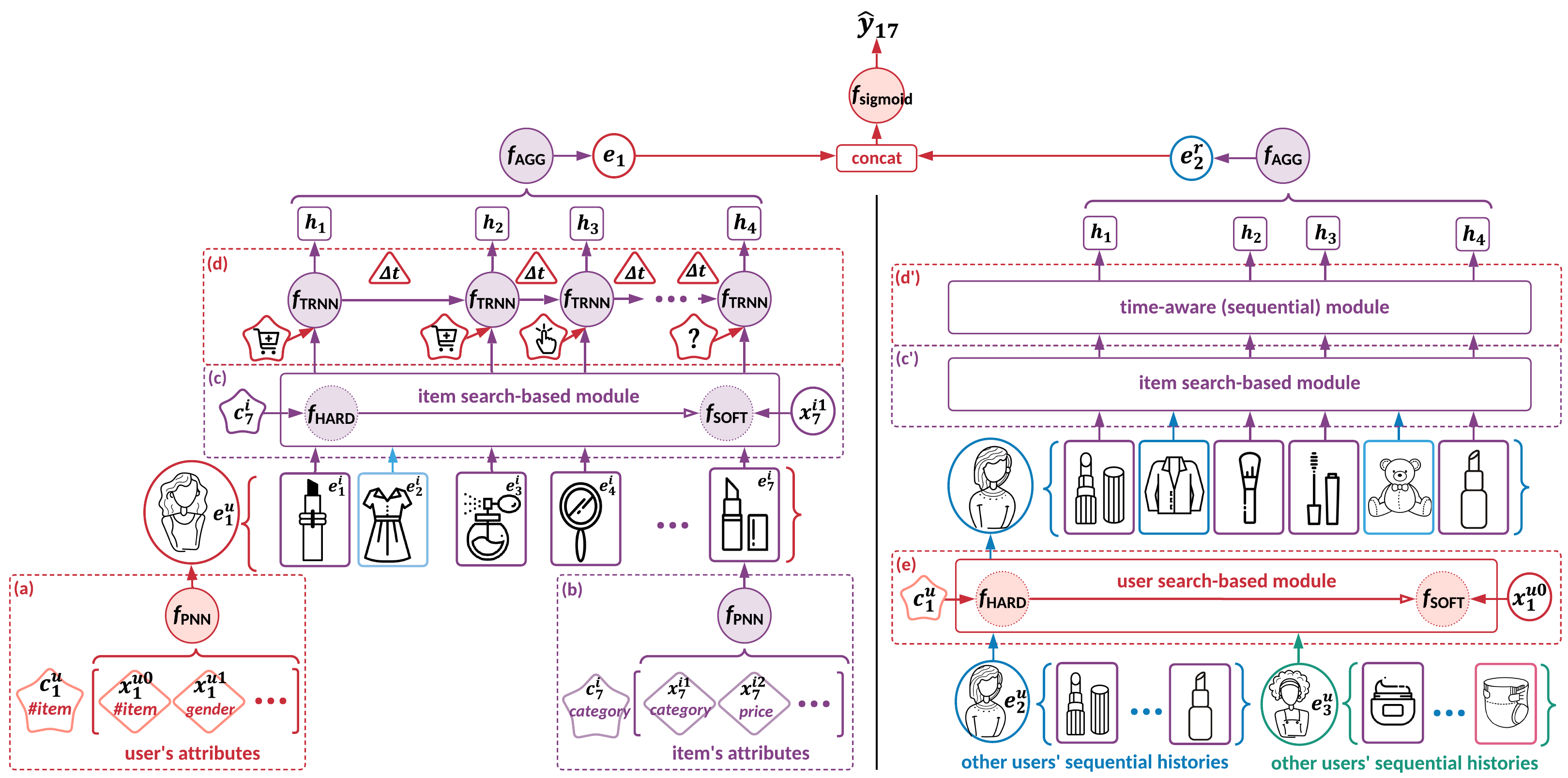}
	\vspace{-5mm}
	\caption{
		The overview of STARec.
		In (a)(b), we use PNN to encode the categorical attributes for both users and items, if available. 
		Notably, $\bm{x}_i^{u0}$ is a manually created feature that denotes the embedding vector of $c^u_1$, and $c^u_1$ is the number of items in user $u_1$'s browsed histories sharing the same category with the target item $i_7$.
		In (c)(d), for each user-item pair, we construct an adaptive search-based module to select relevant items from the whole browsing logs and then feed them into a time-aware (sequential) module.
		Moreover, in (e), we regard the browsing histories from similar users as the additional reference to assist the final prediction (i.e., $\widehat{y}_{17}$ for user $u_1$ and item $i_7$) making. 
		We illustrate the proposed label trick in (d), where previous user feedbacks are used as the input to recover the label of the current item.
	}
	\label{fig:overview}
	\vspace{-2mm}
\end{figure*}

\section{Preliminaries}
\subsection{Related Work}
\minisection{Search-based Recommendation Model}
Classical recommendation methods are proposed to recommend desired items to users based on rich user-item interaction histories either in tabular format \citep{rendle2010factorization,koren2009matrix,guo2017deepfm,juan2016field}, sequence structure \citep{zhou2019deep,zhou2018deep,bai2019ctrec}, or graph structure \citep{jin2020efficient,wang2019neural}.
However, as stated in \citep{qin2020user}, since the users are accumulating more and more behavioral data nowadays, it's non-trivial to train the model from the whole user logs due to the limitations from the online computations.
One feasible solution is only to focus on user's recent records and generate personalized recommendations based on a short sequence instead of a long history \citep{hidasi2015session,tan2016improved,quadrana2017personalizing,ruocco2017inter,donkers2017sequential,chatzis2017recurrent,pei2017interacting}.
However, as recently proposed works \citep{qin2020user,pi2020search} suggest, these methods are not able to encode the periodicity or long-term dependency, which leads to sub-optimal solutions.
Based on this observation, \citet{qin2020user,pi2020search} further propose to build a search model either following a hard-search or soft-search strategy over the whole behavioral history.
In this way, they can use those relevant items instead of the whole set of user-browsed items to efficiently learn a recommendation method.  
Unfortunately, these existing search-based methods overlook effects from time intervals among user's behaviors and thus can not fully use user's browsing sequence.

\minisection{Time-aware (Sequential) Model}
Recent studies \citep{hosseini2018recurrent,zhu2017next} have paid attention on leveraging the time intervals among user's behaviors to better capture user's preferences, for which traditional sequential architectures \citep{hidasi2015session,beutel2018latent,hidasi2018recurrent,jing2017neural,liu2016context} are insufficient.
One direction \citep{zhao2018go,liu2016unified,zhu2017next} is to develop the specific time gates to control the short-term and long-term interest updates.
For example, \citet{zhao2018go} introduces a distance gate based on a recurrent network to control the short-term and long-term point-of-interest updates.
Another way \citep{dai2016deep,hosseini2018recurrent,mei2016neural,vassoy2019time,bai2019ctrec} to integrate time interval information is to formulate the user's sequential histories by a point process, in which the discrete events in user's histories can be modeled in continuous time. 
For example, \citet{mei2016neural} proposes a neural hawkes process model which allows past events to influence the future prediction in a complex and realistic fashion.
We argue that despite computation cost and time complexity, directly feeding long sequential user behavioral data into these methods will bring much more noise which makes it nearly impractical to capture the rich sequential patterns in the user logs.

In this paper, we combine the advantages from both search-based and time-aware models to efficiently retrieve relevant items and mine sequential patterns in an end-to-end way.
Our paper is also related to the label trick proposed in \citep{wang2021bag,shi2020masked} based graph structure.
Instead, our work focuses on the label usage in the sequence cases, which, notably, is also different from the masking technique in existing sequential models such as BERT \citep{devlin2018bert} performing on the feature dimension instead of the label dimension.

\subsection{Problem Formulation}
We begin by clarifying the recommendation task we study and introducing associated notations we use.
\begin{definition}
\label{def:recommendation}
\textbf{Search-based Time-aware Recommendation}\footnote{Notably, a more precise way for notations is to use them in the continuous setting, e.g., define $\mathcal{H}_u$ as $\mathcal{H}_u=$ $\{i_{t_1}$,$i_{t_2}$,$\ldots$,$i_{t_T}\}$ where $t_1,t_2,\ldots,t_T$ denote continuous timestamps, and replace $T+1$ by $T+\epsilon$.}. 
Given a tuple $\langle \mathcal{U}, \mathcal{I}, \mathcal{C}\rangle$ where $\mathcal{U}$ is the set of users, $\mathcal{I}$ is the set of items, $\mathcal{C}$ is the set of items' categories.
For each user $u_m\in\mathcal{U}$, her historical records can be formulated as a sequence of items sorted by time $\mathcal{H}_m=\{i_{1},i_{2},\ldots,i_{T}\}$ where $i_{t}\in \mathcal{H}_m$ is the item browsed by user $u$ at time $t$.
For each item $i_n \in \mathcal{I}$, let $c^i_n$ denote its category (ID).
We use $\bm{x}^u_m$, $\bm{x}^i_n$ to denote the feature of the $m$-th user, the $n$-th item respectively, and further use $\bm{x}^{up}_m$, $\bm{x}^{ip}_n$ to denote their $p$-th categorical features.
The goal of the recommendation is to infer the probability of the user $u_m$ clicking or purchasing the item $i_n$ at a future time $T+1$ conditioning on the retrieved user historical records, denoted as $\widehat{\mathcal{H}}_m=\{\widehat{i}_1,\widehat{i}_2\ldots,\widehat{i}_{\widehat{T}}\}$ where $\widehat{T}$ is the length of retrieval.
\end{definition} 
For convenience, in the following sections, we use the $1$-th categorical feature of each item $i_n$ to represent its category (ID) (e.g., cosmetics).
Namely, $\bm{x}^{i1}_n$ denotes the feature for $n$-th item's category.
For each user $u_m$, we also manually calculate the number of items sharing the same category with target item and further use $c^u_m$ to denote the number.
Note that this number is determined by each user-item pair, not solely depends on user.
Moreover, for each user $u_m$ and item $i_n$ we introduce $\widehat{\mathcal{S}}_m=\{\widehat{u}_1,\widehat{u}_2,\ldots,\widehat{u}_{\widehat{M}}\}$ to denote a set of users similar to $u_m$ being aware of $i_n$.
The computations of $\widehat{\mathcal{H}}_m$ and $\widehat{\mathcal{S}}_m$ are introduced in the following Eqs.~(\ref{eqn:harditem}) and (\ref{eqn:harduser}).

As discussed in \citep{bai2019ctrec}, regardless of search-based part, this time-aware recommendation task (called continuous-time recommendation in \citep{bai2019ctrec}) can be regarded as a generalized sequential recommendation problem of the next-item and next-session/basket problems.
Notably, different from existing solutions to this problem, our method, with the help of the search-based module, can particularly model the time intervals among the relevant items to answer ``\emph{How often does she purchase cosmetics?}'' instead of ``\emph{How often does she purchase items?}''
Easy to see, the answers to the former question are much more informative than the latter one.

\section{The STARec Model}
\subsection{Intuition and Overview}
The basic idea of STARec is to combine the advantages from search-based and time-aware modules, which is based on the following intuitions (as illustrated in Figure~\ref{fig:motivation}):
\begin{itemize}[topsep = 3pt,leftmargin =5pt]
\item \textbf{(I1)} When predicting a user's (e.g., $u_1$) interest in an item (e.g., $i_1$), we need to answer ``\emph{whether $u_1$ wants to buy a lipstick}'' and ``\emph{whether $i_1$ (or its price) is suitable for $u_1$}'', both of which motivate us to search and retrieve the relevant items from her records instead of using the whole browsing items.
\end{itemize}
Specifically, for the first one, although there must be numerous personalized reasons for buying a lipstick, we argue that the popular ones either lie in running out of her old one $i_2$ or wanting an accessory for her new purchases (e.g., lip gloss $i_4$), but not likely due to her purchased iPhone $i_3$.
Also, as for the second one, the prices of these relevant items in $u_1$'s browsing history (e.g., her previous lipstick $i_2$) can give us a useful hint for the suitable price of lipstick in her mind while those irrelevant ones (e.g., her purchased iPhone $i_3$) are much less informative.
\begin{itemize}[topsep = 3pt,leftmargin =5pt]
\item \textbf{(I2)} User's interests are naturally diverse and always drifting, which can be captured from their behaviors.
However, each interest has its own evolving process. 
For example, a teen may purchase lipsticks weekly, and phones yearly, and purchasing lipsticks only has a slight effect on purchasing phones.
It supports us to build a time-aware module for each class of items.
\item \textbf{(I3)} User's current behavior can be significantly influenced by her previous ones.
For example, a user is likely to stop browsing after clicking or purchasing an item since she has already found her favorite.
It motivates us to include user feedbacks (i.e., labels) as the input.
\end{itemize}







Figure~\ref{fig:overview} shows the overview of STARec.
First, we use a product-based neural network (PNN) \citep{qu2016product} to model the correlations of categorical features (if available) for each user and item, as shown in (a)(b).
After that, we develop a novel adaptive search-based module to retrieve relevant items based on the similarities between their categories and the target item's category, and then use a time-aware module to mine their sequential patterns, as (c)(d) illustrate.
Moreover, we also retrieve those similar users' histories and regard this information as the additional references to assist the final prediction making, as (e) shows.
Note that besides this architecture, we propose to involve the user's previous feedbacks (i.e., labels) in the input, as illustrated in (d).


\subsection{Object Modeling}

If we are not accessible to the rich categorical features for each user and item, we can build an embedding vector (i.e., $\bm{e}^u_m$ and $\bm{e}^i_n$) for each user (ID) $u_m$ and item (ID) $i_n$.
Otherwise, we need to consider rich correlations among these features, which play an important role in user interest modeling.
For instance, in Figure~\ref{fig:overview}, $u_1$ is likely to purchase $i_1$ because she wants to buy a lipstick \textbf{AND} its price is suitable for her.
As discussed in \citep{rendle2010factorization,qu2016product}, this ``AND'' operation can not be solely modeled by classical neural network (e.g., multi-layer perceptron (MLP)) but can be captured by the product-based neural network (PNN) model \citep{qu2016product}.
Therefore, we adopt PNN to capture the hidden correlations among the categorical features for each user and item.
Specifically, its output of $n$-th item can be formulated as 
\begin{equation}
\label{eqn:fm}
\bm{e}^i_n \coloneqq f^i_\mathtt{PNN}(\bm{x}^i_n) =  \bm{v}^i_n \odot \bm{x}^i_n + \sum^P_{p=1}\sum^P_{p'=p+1} (\bm{v}^i_{p} \odot \bm{v}^i_{p'}) \bm{x}^{ip}_n \bm{x}^{ip'}_n,
\end{equation}
where $\odot$ denotes the element-wise product operation and $\bm{v}^i_n$, $\bm{v}^i_p$ denote learnable weights.
In Eq.~(\ref{eqn:fm}), the first term shows the first-order feature interactions, and the second term illustrates the second-order feature interactions.
As for each user, similarly, we can define $\bm{e}^u_m\coloneqq f^u_\mathtt{PNN}(\bm{x}^u_m)$ as the output of the $m$-th user where $f^u_\mathtt{PNN}(\cdot)$ and $f^i_\mathtt{PNN}(\cdot)$ share the same formulation but with different parameters.

\subsection{Search-based Module}
\minisection{Item Search-based Module}
As discussed in \citep{pi2020search}, the categories are one of the most powerful tools to measure the similarity (i.e., relevance) between the user's browsed items and target item.
Based on this, we can easily derive a hard-search strategy.
Formally, we first construct a set of items for each user $u_m$ defined as
\begin{equation} 
\label{eqn:harditem}
\widehat{\mathcal{H}}_m \coloneqq \{i_{n'}| i_{n'}\in\mathcal{H}_m \land f^i_\mathtt{HARD}(c^i_{n'},c^i_{n})\geq\epsilon\}\cup \mathcal{H}^\mathtt{RECENT}_m,
\end{equation}
where except $\mathcal{H}^\mathtt{RECENT}_m$ denotes a set of recent browsed items, the first term (i.e., $i_{n'}\in\mathcal{H}_m$) limits the retrieved items to come from $u_m$'s browsing history and the second term (i.e., $f^i_\mathtt{HARD}(c^i_{n'},c^i_{n})\geq\epsilon$) selects the relevant ones, $\epsilon\in[0,1]$ denotes the threshold,
$c_{n'},c_{n}$ are one-hot vectors directly represent their categories without any learnable parameter.
In this case, $f^i_\mathtt{HARD}(\cdot, \cdot)$ can be defined as $f^i_\mathtt{HARD}(c^i_{n'},c^i_{n})\coloneqq-|c^i_{n'}-c^i_{n}|$ and $\epsilon=0$.

The computation cost for this hard-search strategy is $|\mathcal{H}_m|$ for each user $u_m$ and item $i_n$.
It is very efficient, but sometimes too hard-and-fast.  
In other words, it can only find those items exactly sharing the same category with the target item.
It doesn't work in the following case where \emph{a teen purchases a beautiful dress and she needs lipstick to make up}.
To handle these cases, we further introduce $f^i_\mathtt{SOFT}(\cdot,\cdot)$ defined as $f^i_\mathtt{SOFT}(\bm{x}^{i1}_{n'},\bm{x}^{i1}_{n})\coloneqq\text{cos}(\bm{x}^{i1}_{n'},\bm{x}^{i1}_{n})=({\bm{x}^{i1}_{n'}}^\top\cdot\bm{x}^{i1}_{n})/(|\bm{x}^{i1}_{n'}|\cdot|\bm{x}^{i1}_{n}|)$ where $\text{cos}(\cdot,\cdot)$ denotes cosine similarity.

One can obtain retrieved items by the soft-search strategy through replacing $f^i_\mathtt{HARD}(c^i_{n'},c^i_{n})$ by $f^i_\mathtt{SOFT}(\bm{x}^{i1}_{n'},\bm{x}^{i1}_{n})$ and assigning $0< \epsilon <1$ in Eq.~(\ref{eqn:harditem}).
In this case, the performance of the soft-search largely depends on how well learnable vectors $\bm{x}^{i1}_{n'}$, $\bm{x}^{i1}_{n}$ are trained. 
Existing methods either introduce a mixed loss function \citep{pi2020search} or apply a reinforcement learning algorithm \citep{qin2020user} in training.
Instead, we propose a simpler and more effective way: an adaptive search strategy, which combines the advantages from both the hard-search and soft-search strategies and also enables the whole architecture to be trained in an end-to-end fashion. 

We first employ a sign function denoted as $\mathtt{sgn}(\cdot)$ to re-scale the hard-search and use a softmax function denoted as $f_\mathtt{softmax}(\cdot,\cdot)$ to re-scale the soft-search.
Formally, we define our adaptive search strategy $f^i_\mathtt{ADA}(c^i_{n'},c^i_{n},\bm{x}^{i1}_{n'},\bm{x}^{i1}_{n})$ as
\begin{equation}
\begin{aligned}
\label{eqn:searchitem}
f^i_\mathtt{ADA}&(c^i_{n'},c^i_{n},\bm{x}^{i1}_{n'},\bm{x}^{i1}_{n}) \coloneqq
-\frac{\mathtt{sgn}(|c^i_{n'}-c^i_{n}|)}{1-\tau} + f_\mathtt{softmax}(\text{cos} (\bm{x}^{i1}_{n'},\bm{x}^{i1}_{n}), \tau)
\\
&=-\frac{\mathtt{sgn}(|c^i_{n'}-c^i_{n}|)}{1-\tau}+\frac{\text{exp}(\text{cos}(\bm{x}^{i1}_{n'},\bm{x}^{i1}_{n} )/\tau)}{\sum_{i_{n''}\in\mathcal{H}_m}\text{exp}(\text{cos}(\bm{x}^{i1}_{n''},\bm{x}^{i1}_{n})/\tau)},
\end{aligned}
\end{equation}
where $\tau\in(0,1)$ denotes temperature hyper-parameter to balance the hard-search and the soft-search.
In practice, we set the initial temperature as $0.99$, then gradually reduce the temperature until $0.01$ during the training process.
One can see that at the beginning, the first term (i.e., hard-search) plays a major part in the search, and as the training goes on, the second term (i.e., soft-search) is playing a more and more important role in the search.
Therefore, with the help of this adaptive search, our whole architecture is able to be trained in an end-to-end fashion.

Besides using those relevant items from the item aspect, we also consider including similar experiences from relevant users as the reference.
The motivation behind this is very straightforward that, as shown in Figure~\ref{fig:motivation}(b), the teens $u_1$, $u_2$ often share similar interests over items and similar browsing patterns, which are usually different from a young mother $u_3$.
Hence, the browsing records of $u_1$, $u_2$ would benefit each other, and that of $u_3$ would be noise when modeling the browsing patterns of $u_1$.

Based on this observation, besides the item search-based module introduced above, we further construct a user search-based module, whose target is to find similar users for further including their records as the references to help with the final prediction making.
Formally, for each user $u_m$ and item $i_n$, we construct a set of retrieved users similar with $u_m$ being aware of $i_n$ as
\begin{equation}
\label{eqn:harduser}
\widehat{\mathcal{S}}_m \coloneqq \{u_{m'}|u_{m'}\in\mathcal{U}\land f^u_\mathtt{HARD}(c^u_m,c^u_{m'})\geq \eta\},
\end{equation}
where analogous to Eq.~(\ref{eqn:harditem}), $c_{m'},c_{m}$ are one-hot vectors directly representing the numbers of items in their histories sharing the same category with $i_n$ without any learnable parameters, and $f^u_\mathtt{HARD}(\cdot, \cdot)$ can be defined as $f^u_\mathtt{HARD}(c^u_{m'},c^u_{m})\coloneqq-|c^u_{m'}-c^u_{m}|$ and $\eta$ is a threshold.
Similarly, we define $f^u_\mathtt{SOFT}(\cdot,\cdot)$ as $f^u_\mathtt{SOFT}(\bm{e}^{u}_{m'},\bm{e}^{u}_{m})\coloneqq \text{cos}(\bm{e}^{u}_{m'},\bm{e}^{u}_{m})$,
and propose an adaptive search strategy $f^u_\mathtt{ADA}(c^u_{m'},c^u_{m},\bm{e}^{u}_{m'},\bm{e}^{u}_{m})$ from user aspect as
\begin{equation}
\begin{aligned}
\label{eqn:searchuser}
f^u_\mathtt{ADA}&(c^u_{m'},c^u_{m},\bm{e}^{u}_{m'},\bm{e}^{u}_{m}) \coloneqq
-\frac{\mathtt{sgn}(|c^u_{m'}-c^u_{m}|)}{1-\iota} + f_\mathtt{softmax}(\text{cos}( \bm{e}^{u}_{m'},\bm{e}^{u}_{m}), \iota)
\\
&=-\frac{\mathtt{sgn}(|c^u_{m'}-c^u_{m}|)}{1-\iota}+\frac{\text{exp}(\text{cos}(\bm{e}^{u}_{m'},\bm{e}^{u}_{m} )/\iota)}{\sum_{u_{m''}\in\mathcal{U}}\text{exp}(\text{cos}(\bm{e}^{u}_{m''},\bm{e}^{u}_{m})/\iota)},
\end{aligned}
\end{equation}
where $\iota\in(0,1)$ denotes temperature hyper-parameter to balance the hard and soft parts.
After we obtain the set of similar users $\widehat{\mathcal{S}}_m$ for each user $u_m$, we then employ the item search-based module to construct a set of user browsing histories $\{\widehat{\mathcal{H}}_{m'}|u_{m'}\in\widehat{\mathcal{S}}_m\}$.

\subsection{Time-aware Module}
Given a user $u_m$ and an item $i_n$, we have a set of retrieved items in $u_m$'s browsing history $\widehat{\mathcal{H}}_m$ and a set of similar users' browsing histories $\{\widehat{\mathcal{H}}_{m'}|u_{m'}\in\widehat{\mathcal{S}}_m\}$.

\begin{figure}[t]
	\centering
	\vspace{-2mm}
	\includegraphics[width=1.0\linewidth]{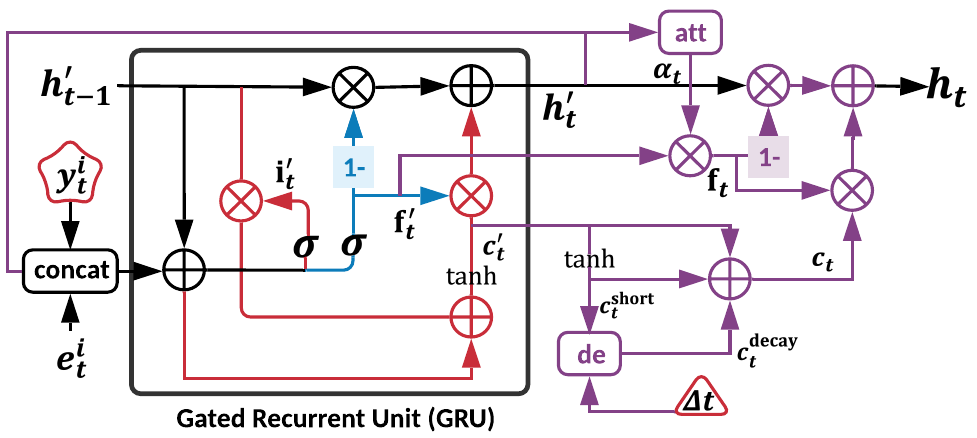}
	\vspace{-6mm}
	\caption{
		An illustrated example for time-aware module denoted as $f_\mathtt{TRNN}$ in Figure~\ref{fig:overview}, where the input of the $t$-th cell is the concatenation of item's original feature $\bm{x}^i_t$ and embedding vector of label $\bm{y}^i_t$.
		We incorporate modified GRU with an attention mechanism to model the user's sequential pattern, considering the effect of her previous feedbacks and time intervals in browsing history.
	}
	\label{fig:seq}
	\vspace{-5mm}
\end{figure}

For each $i_n\in\widehat{\mathcal{H}}_m$, we use $y_n$ to denote the feedback from user $u_m$.
It's straightforward to correspondingly build a one-hot vector or an embedding vector $\bm{y}^i_n$ here.
Hence, as illustrated in Figure~\ref{fig:seq}, the input of the $t$-th cell of our time-aware module (denoted as $f_\mathtt{TRNN}$) is the concatenation of item's original feature and embedding vector of label (denoted as $[\bm{e}^i_t,\bm{y}^i_t]$).

In order to empower our time-aware module for modeling the user shifting interest over time, as shown in Figure~\ref{fig:seq}, we first adopt a gated recurrent unit (GRU) to mine the useful sequential patterns, which can be formulated as
\begin{equation}
\begin{aligned}
\label{eqn:lstm}
\text{f}'_t &= \sigma(\bm{W}_{\text{f}'}[\bm{e}^i_t,\bm{y}^i_t]+\bm{U}_{\text{f}'}\bm{h}'_{t-1}
), \ \text{i}'_t =\sigma(\bm{W}_{\text{i}'}[\bm{e}^i_t,\bm{y}^i_t]+\bm{U}_{\text{i}'}\bm{h}'_{t-1}
), \\
\bm{c}'_t &= \mathtt{tanh}(\bm{W}_{c'}[\bm{e}^i_t,\bm{y}^i_t]+\text{i}'_t \odot \bm{U}_{c'}\bm{h}'_{t-1}
), \
\bm{h}'_t = \text{f}'_t \odot \bm{c}'_{t} +  (1-\text{f}'_t)\odot \bm{h}'_{t-1},
\end{aligned}
\end{equation}
where we omit the bias term for simplicity.
We further use attention mechanism to model the evolution of user interest and consider the effect from time intervals as
\begin{equation}
\begin{aligned}
\label{eqn:attention}
&\bm{c}^\mathtt{short}_t = \mathtt{tanh}(\bm{W}_c\bm{c}'_t+\bm{b}_c), \ \bm{c}^\mathtt{decay}_t = \bm{c}^\mathtt{short}_t \cdot \mathtt{de}(\Delta t),\\
&\alpha'_t =  \text{exp}(\bm{h}'_t\bm{W}_{\alpha'}[\bm{e}^i_t,\bm{y}^i_t])/\sum^{\widehat{T}+1}_{t'=1}(\bm{h}'_{t}\bm{W}_{\alpha'}[\bm{e}^i_{t'},\bm{y}^i_{t'}]), \ \text{f}_t = \alpha'_t \cdot \text{f}'_t,\\
&\bm{c}_t = \bm{c}'_t - \bm{c}^\mathtt{short}_t + \bm{c}^\mathtt{decay}_t, \ \bm{h}_t =  \text{f}_t \odot \bm{c}_{t} +  (1-\text{f}_t)\odot \bm{h}'_{t},
\end{aligned}
\end{equation}
where $\Delta t$ is the elapsed time between items $i_{t-1}$ and $i_t$, and $\mathtt{de}(\cdot)$ denotes a heuristic decaying function.
We use $\mathtt{de}(\Delta t) = 1/ \Delta t$ for datasets with small amount of elapsed time and $\text{de}(\Delta t) = 1/ \text{log}(e+\Delta t)$ for those with large elapsed time in practice.

As a consequence, for each sequence $\widehat{\mathcal{H}}_m$ or one in $\{\widehat{\mathcal{H}}_{m'}|u_{m'}\in\widehat{\mathcal{S}}_m\}$, we obtain a set of hidden states $\{\bm{h}_t|i_t\in\widehat{\mathcal{H}}_m\}$ or $\{\bm{h}_t|i_t\in\widehat{\mathcal{H}}_{m'}\}$.
For each set of hidden states, we employ an aggregation function $f_\mathtt{AGG}(\cdot)$ to fuse these embeddings into the representation of the whole sequence,
which, taking $\{\bm{h}_t|i_t\in\widehat{\mathcal{H}}_m\}$ as an instance, can be formulated as
\begin{equation}
\label{eqn:aggregation}
\bm{e}_m = f_\mathtt{AGG}(\{\bm{h}_t|i_t\in\widehat{\mathcal{H}}_m\}) = \sigma(\bm{W}_m\cdot(\sum_{i_t\in\widehat{\mathcal{H}}_m}\alpha_t(\bm{h}_t\bm{W}_t))+\bm{b}_m),
\end{equation}
where $\alpha_t = \text{exp}(\bm{h}_t\bm{W}_\alpha)/\sum^{\widehat{T}+1}_{t'=1}(\bm{h}_{t}\bm{W}_\alpha)$.
Similarly, we can obtain $\{\bm{e}^r_{m'}|u_{m'}\in\widehat{\mathcal{S}}_m\}$
where $\bm{e}^r_{m'} = f_\mathtt{AGG}(\{\bm{h}_t|i_t \in \mathcal{H}_{m'}\})$ for sequence $\widehat{\mathcal{H}}_{m'}$ in  $\{\widehat{\mathcal{H}}_{m'}|u_{m'}\in\widehat{\mathcal{S}}_m\}$.

Notably, as introduced in Eq.~(\ref{eqn:harditem}),  $\widehat{\mathcal{H}}_m$ for each user $u_m$ consists of two parts: one is a set of the recent browsed items (i.e., $\widehat{\mathcal{H}}^\mathtt{RECENT}_m$), the other is a set of retrieved items (i.e., $\widehat{\mathcal{H}}_m /\widehat{\mathcal{H}}^\mathtt{RECENT}_m$).
In the implementation, we establish two sequential networks (without parameter sharing).
We use one sequential network for each part to encode these items and then combine these outputs together by concatenation.
We demonstrate that this way is more efficient than putting all the items in one sequential network in Section~\ref{sec:result}.  

\subsection{Optimization Objective}
For each user-item pair $(u_m,i_n)$, we generate the final prediction $\widehat{y}_{mn}$ by encoding $\bm{e}_m$ and $\{\bm{e}^r_{m'}|u_{m'}\in\widehat{\mathcal{S}}_m\}$.
Specifically, we combine a sigmoid function with a MLP layer over the concatenation of these embeddings as
\begin{equation}
\label{eqn:prediction}
\widehat{y}_{mn} = \mathtt{sigmoid}(f_\mathtt{MLP}([\bm{e}_m,\{\bm{e}^r_{m'}|u_{m'}\in\widehat{\mathcal{S}}_m\}])).
\end{equation}
After that, we adopt a log loss to update the parameter $\theta$ as
\begin{equation}
\label{eqn:loss}
\mathcal{L}_\theta = - \sum_{(u_m,i_n)\in\mathcal{D}} (y_{mn}\cdot \text{log}(\widehat{y}_{mn})+(1-y_{mn})\cdot \text{log}(1-\widehat{y}_{mn})),
\end{equation}
where $\mathcal{D}$ denotes the datasets containing the true label $y_{mn}$ for each user-item pair $(u_m, i_n)$.
We provide a detailed pseudo code of the training process and the corresponding time complexity analysis in Appendix~\ref{app:algorithm}.

\section{Experiments}
\label{sec:experiment}
\subsection{Dataset and Experimental Flow}
We use three large-scale real-world datasets, namely Tmall\footnote{\url{https://tianchi.aliyun.com/dataset/dataDetail?dataId=42}}, Taobao\footnote{\url{https://tianchi.aliyun.com/dataset/dataDetail?dataId=649}}, Alipay\footnote{\url{https://tianchi.aliyun.com/dataset/dataDetail?dataId=53}}, which contain users online records from three corresponding platforms of Alibaba Group.
Please refer to Appendix~\ref{app:dataset} for detailed description of the datasets and \ref{app:config} for detailed experimental configuration.

\subsection{Baselines and Evaluation Metrics}
We compare our model mainly against 13 representative recommendation methods including LSTM \citep{hochreiter1997long}, RRN \citep{wu2017recurrent}, STAMP \citep{liu2018stamp}, Time-LSTM \citep{zhu2017next}, NHP \citep{mei2016neural}, DUPN \citep{ni2018perceive}, NARM \citep{li2017neural}, ESMM \citep{ma2018entire}, ESM$^2$ \citep{wen2019entire}, MMoE \citep{ma2018modeling}, DIN \citep{zhou2018deep}, DIEN \citep{zhou2019deep}, SIM \citep{pi2020search}.
In order to further investigate the effect from each component of STARec, we design the following three variants:
\begin{itemize}[topsep = 3pt,leftmargin =5pt]
\item \textbf{STARec} is our model without using user previous feedbacks for fair comparsion.
\item \textbf{STARec}$^-_\mathtt{time}$ is a variant of STARec using a standard LSTM as the time-aware (sequential) module.
\item \textbf{STARec}$^-_\mathtt{recent}$ is a variant of STARec where $\mathcal{H}^\mathtt{RECENT}_m$ is not included in $\widehat{\mathcal{H}}_m$ (see Eq.~(\ref{eqn:harditem})).
\item \textbf{STARec}$^+_\mathtt{label}$ is a variant of STARec using user's previous feedbacks as input.
\end{itemize}
We provide the descriptions of these baseline methods in Appendix~\ref{app:baseline}. 
We provide detailed descriptions of experimental settings and data pre-processing in Appendix~\ref{app:config}.
To evaluate the performance of the above methods, we choose Area under the ROC Curve (AUC), Accuracy (ACC), LogLoss as the evaluation metric.
The thresholds of ACC on Tmall and Alipay datasets are set as 0.5, while that on the Taobao dataset is set as 0.1 due to a large number of negative instances. 

\begin{table*}[t]
	\centering
	\caption{Comparison of different (sequential) recommendation models on three industrial datasets. Results of Click-Through Rate (CTR) prediction task are reported. 
	* indicates $p < 0.001$ in significance tests compared to the best baseline. 
	Note that our results are not consistent with the results in \citep{qin2020user} due to different experimental settings. Refer to details in Appendix~\ref{app:config}.}
	\vspace{-3mm}
	\resizebox{0.82\textwidth}{!}{
		\begin{tabular}{@{\extracolsep{4pt}}cccccccccc}
			\toprule
			\multirow{2}{*}{Recommender} & \multicolumn{3}{c}{Tmall} & \multicolumn{3}{c}{Alipay} & \multicolumn{3}{c}{Taobao} \\
			\cmidrule{2-4}
			\cmidrule{5-7}
			\cmidrule{8-10}
			{} & AUC & ACC & LogLoss & AUC & ACC & LogLoss & AUC & ACC & LogLoss \\
			\midrule
			LSTM & 
			0.6973 & 0.7054 & 0.5854 & 
			0.8357 & 0.7697 & 0.4713 & 
			0.5912 & 0.4516 & 0.4411 \\ 
			\midrule
			\textbf{LSTM}$^+_\mathtt{label}$ & 
			\textbf{0.7662} & \textbf{0.7324} & \textbf{0.5291} & 
			\textbf{0.9052} & \textbf{0.8449} & \textbf{0.3738} & 
			\textbf{0.6450} & \textbf{0.5780} & \textbf{0.4094} \\ 
			\midrule
			RRN & 
			0.6973 & 0.7073 & 0.5866 & 
			0.8429 & 0.7145 & 0.4636 & 
			0.5102 & 0.1502 & 0.4398 \\
			\midrule
			Time-LSTM & 
			0.6962 & 0.6796 & 0.5865 & 
			0.8439 & 0.8057 & 0.4874 &
			0.5945 & 0.4393 & 0.4397 \\
			\midrule
			NHP & 
			0.6979 & 0.6969 & 0.5849 & 
			0.8490 & 0.7922 & 0.4743 & 
			0.6003 & 0.3205 & 0.4393 \\
			\midrule
			DUPN & 
			0.5551 & 0.6796 & 0.6269 & 
			0.8021 & 0.7552 & 0.5243 & 
			0.5525 & 0.1921 & 0.4471 \\
			\midrule
			NARM & 
			0.6396 & 0.6839 & 0.6025 & 
			0.8422 & 0.7695 & 0.4860 & 
			0.5972 & 0.3426 & 0.4494 \\
			\midrule
			STAMP & 
			0.6753 & 0.6946 & 0.5829 & 
			0.8178 & 0.7491 & 0.5137 & 
			0.6012 & 0.3592 & 0.4448 \\
			\midrule
            ESMM & 
			0.5189 & 0.6796 & 0.6312 & 
			0.7131 & 0.6856 & 0.6080 &
			0.5487 & 0.1774 & 0.4573 \\
			\midrule
			ESM$^2$ & 
			0.5149 & 0.6796 & 0.6310 & 
			0.7241 & 0.6930 & 0.5996 & 
			0.5030 & 0.1520 & 0.4594 \\
			\midrule
			MMoE & 
			0.5060 & 0.6796 & 0.6313 & 
			0.7119 & 0.6834 & 0.6085 & 
			0.5501 & 0.1719 & 0.4565 \\
			\midrule
			DIN & 
			0.6878 & 0.6946 & 0.5915 & 
			0.8496 & 0.7692 & 0.4717 &
			0.5978 & 0.4422 & 0.4388 \\
			\midrule
			DIEN & 
			0.6892 & 0.6962 & 0.5833 & 
			0.8474 & 0.7949 & 0.4668 & 
			0.5963 & 0.4025 & 0.4412 \\
			\midrule
			SIM & 
			0.7005 & 0.7094 & 0.5698 & 
			0.8549 & 0.8069 & 0.4623 &
			0.6045 & 0.4538 & 0.4271 \\
			\midrule
			STARec$^-_\mathtt{time}$ & 
			0.6999 & 0.7097 & 0.5684 & 
			0.8527 & 0.8046 & 0.4547 & 
			0.6035 & 0.4525 & 0.4312 \\
			\midrule
			STARec$^-_\mathtt{recent}$ & 
			0.7013 & 0.7081 & 0.5632 & 
			0.8536 & 0.8012 & 0.4672 & 
			0.6021 & 0.4561 & 0.4355 \\
			\midrule
			\textbf{STARec} & 
			\textbf{0.7204}$^*$ & \textbf{0.7150}$^*$ & \textbf{0.5471}$^*$ & 
			\textbf{0.8624}$^*$ & \textbf{0.8142}$^*$ & \textbf{0.4410}$^*$ &
			\textbf{0.6126}$^*$ & \textbf{0.4629}$^*$ & \textbf{0.4211}$^*$ \\
			\midrule
			\textbf{STARec}$^+_\mathtt{label}$ & 
		    \textbf{0.7986}$^*$ & \textbf{0.7502}$^*$ & \textbf{0.5059}$^*$ & 
			\textbf{0.9201}$^*$ & \textbf{0.8661}$^*$ & \textbf{0.3423}$^*$ & 
			\textbf{0.6771}$^*$ & \textbf{0.6039}$^*$ & \textbf{0.3862}$^*$ \\
			\bottomrule
		\end{tabular}
	}
	\label{tab:res}
	\vspace{-2mm}
\end{table*}	

\begin{figure}[h]
	\centering
	\includegraphics[width=1.0\linewidth]{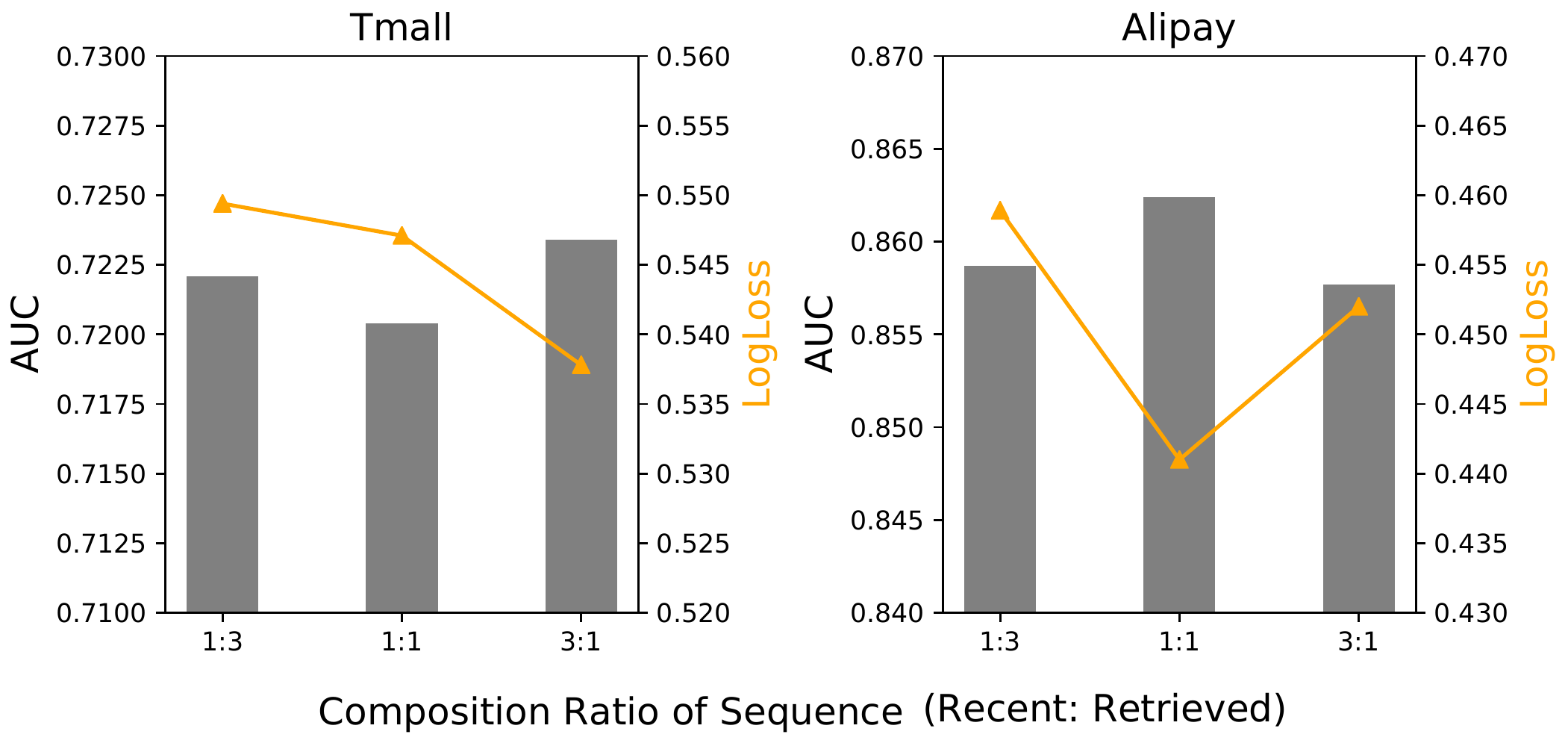}
	\vspace{-7mm}
	\caption {Comparison of performance of STARec under different composition ratios of recent and retrieved items in sequence on Tmall and Alipay datasets, in terms of AUC and LogLoss.}
	\label{fig:recent}
	\vspace{-5mm}
\end{figure}

\subsection{Result Analysis}
\label{sec:result}
\minisection{Overall Performance}
Table~\ref{tab:res} summarizes the results.
The major findings from our offline experiments are summarized as follows.
\begin{itemize}[topsep = 3pt,leftmargin =5pt]
\item Our model outperforms all these baseline methods including sequential models (e.g., RRN, LSTM, NHP) and tower architecture based mdels (e.g., ESMM, MMoE, ESM$^2$).
These results may be explained as our model, unlike these methods, combining the advantages of both search-based and time-aware (sequential) models.
\item Compared to those other models (e.g., ESMM, MMoE, ESM$^2$), most of the sequential recommendation methods (e.g., RRN, LSTM, NHP) achieve better performance. 
We may conclude that encoding the contextual information in the historical sequences is crucial to capture user patterns, as whether a user has already found the desired items or not holds a significant effect on user behaviors on the current item.
\item With a comparison between SIM and other existing sophisticated models (e.g., DIN, DIEN), we find that SIM consistently outperforms these methods.
The reason seems to be that SIM introduces a search-based module to use the retrieved relevant information instead of the whole sequences.
\end{itemize}

\begin{figure}[t]
	\centering
	\includegraphics[width=0.8\linewidth]{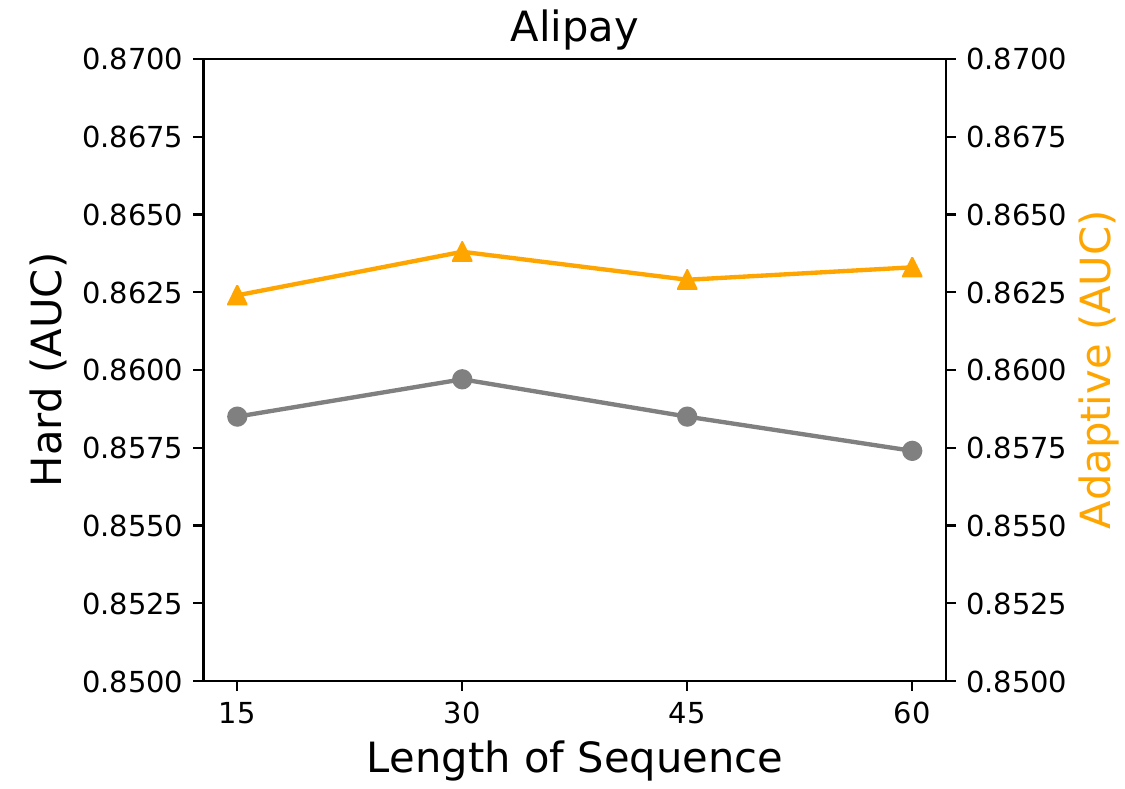}
	\vspace{-2mm}
	\caption{Comparison of performance of STARec with the hard-search or adaptive search strategies under different lengths of sequence on Alipay dataset, in term of AUC.}
	\label{fig:length}
	\vspace{-4mm}
\end{figure}

\minisection{Impact of Recent Histories}
From the comparison between STARec and STARec$^-_\text{recent}$ in Table~\ref{tab:res}, we can observe that replacing some retrieved items with the recent items can consistently improve the performance in all the datasets. 
Specifically, for each dataset, the sequence length of STARec and other baselines is set as 30.
Distinct from other methods, half of sequence of STARec includes retrieved items, while the other half consists of recent ones.
Hence, we here further investigate how the performance of STARec changes when involving more recent ones (and less retrieved ones) or less recent ones (and more retrieved ones).
Figure~\ref{fig:recent} depicts the performance of STARec under three cases.
It's difficult to conclude the best ratio in a general way, as the value varies for different datasets.

\minisection{Impact of Search-based Module}
As Table~\ref{tab:res} shows, we can see that STARec achieves better performance than STARec$^-_\mathtt{search}$ in all these three datasets.
The observation that SIM works better than DIE and DIEN methods also verifies the superiority of search-based models.
As our paper introduces a new adaptive search strategy, we further compare its performance to the hard-search strategy under different sequence lengths.
From Figure~\ref{fig:length}, we see that our proposed adaptive search strategy can consistently outperform the hard-search strategy.
One possible explanation is that the hard-search strategy can be regarded as a special case of our adaptive search strategy.
Also, we observe that their performance gap gets bigger when the length of the sequence reaches 60.
A possible explanation is that the hard-search strategy, at most, only can search and retrieve all the items whose categories are same to the target item, while our adaptive search strategy definitely searches and retrieves items in a larger scope, which can involve more useful information.

\minisection{Impact of Time-aware Module}
In Table~\ref{tab:res}, we compare STARec to STARec$^-_\mathtt{time}$. 
Results show that taking time intervals of user behaviors into consideration can improve the performance of our model, which verifies our idea of building a time-aware module.

\minisection{Impact of Label Trick}
From Table~\ref{tab:res}, one can see that our label trick (using the previous user feedbacks as the input) can significantly improve the performance of STARec.
We further investigate the impact of our label trick with other sequential models (e.g., LSTM).
In Table~\ref{tab:res},  we design LSTM$^+_\mathtt{label}$, a variant of LSTM that uses user previous feedbacks as the input.
Comparison between LSTM and LSTM$^+_\mathtt{label}$ shows the significant improvement from the label trick, which, to an extent, outweights the gains from more dramatic changes in the underlying user modeling architecture.

\begin{figure}[t]
	\centering
 	\includegraphics[width=0.85\linewidth]{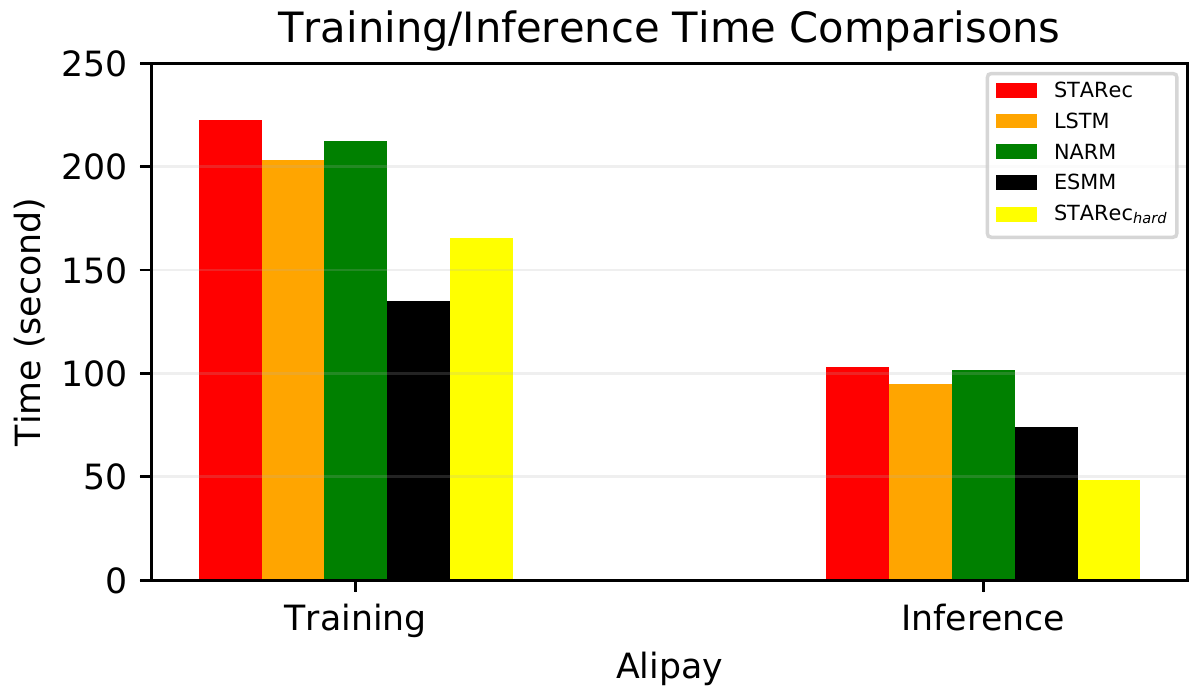}
	\vspace{-4mm}
	\caption {Training/inference time comparisons of STARec and STARec$_\mathtt{hard}$ against baselines on Alipay dataset.}
	\label{fig:time}
	\vspace{-3mm}
\end{figure}

\minisection{Complexity Analysis}
We investigate the time complexity of STARec against baseline methods LSTM, NARM, ESMM, and further introduce STARec$_\mathtt{hard}$ as a variant of STARec using the hard-search strategy.
We then report the training and inference times for one round of the whole data.
From Figure~\ref{fig:time}, we can observe that STARec$_\mathtt{hard}$ is more efficient than STARec, as our adaptive search strategy needs to compute the similarity of category embeddings.
More importantly, we also can see that the training and inference times of STARec$_\mathtt{hard}$ are comparable to, or even smaller than, other baselines.
One explanation is that we employ two sequential networks to model the recent items and retrieved items in STARec and STARec$_\mathtt{hard}$.
Hence, the length of our time-aware module is half of the length of these baselines leading to an efficient implementation.

\section{Real-world Deployment}
\label{sec:realworld}
In order to verify the effectiveness of STARec in real-world applications, we deploy our method in two main item recommendation scenarios (called ``Guess You Like'' and ``Information Flow'') in Company X, a main-stream bank company.
This App has millions of daily active users who create billions of user logs every day in the form of implicit feedbacks such as click behavior.
Please refer to discussions on deployment in Appendix~\ref{app:deploy}.

\begin{table}[h]
	\centering
	\vspace{-2mm}
	\caption{Comparison of different (sequential) recommendation models on real-world recommendation scenarios.}
	\vspace{-3mm}
	\resizebox{0.85\linewidth}{!}{
		\begin{tabular}{@{\extracolsep{4pt}}cccccccccc}
			\toprule
			\multirow{2}{*}{Recommender} & \multicolumn{2}{c}{Guess You Like} & \multicolumn{2}{c}{Information Flow} \\
			\cmidrule{2-3}
			\cmidrule{4-5}
			{} & AUC & CTR & AUC & CTR\\
			\midrule
			DIN & 0.7828 & 1.47\% & 0.8409 & 1.26\%\\ 
			\midrule
			DIEN & 0.8068 & 1.52\% & 0.8493 & 1.30\%\\
			\midrule
			\textbf{STARec} & 
			\textbf{0.8909} & \textbf{1.61\%} & \textbf{0.9159} & \textbf{1.32\%}\\
			\bottomrule
		\end{tabular}
	}
	\label{tab:deploy}
	\vspace{-3mm}
\end{table}

\subsection{Offline Evaluation}
For the offline experiment, we use a daily updated dataset collected from September 11th, 2021 to October 10th, 2021 for training and evaluation.
Concretely, for each user, we use the last 31 user behaviors as the test set and use the rest records as the training set.
The task is CTR prediction, and the overall performance is shown in Table~\ref{tab:deploy}.
From Table~\ref{tab:deploy}, we see that STARec achieves 13.8\% and 8.9\% improvement over DIN \citep{zhou2018deep}, 10.7\% and 7.8\% improvement over DIEN \citep{zhou2019deep} on AUC metric in ``Guess You Like'' and ``Information Flow'' scenarios respectively.

\subsection{Online Evaluation}
\begin{figure}[t]
	\centering
 	\includegraphics[width=0.85\linewidth]{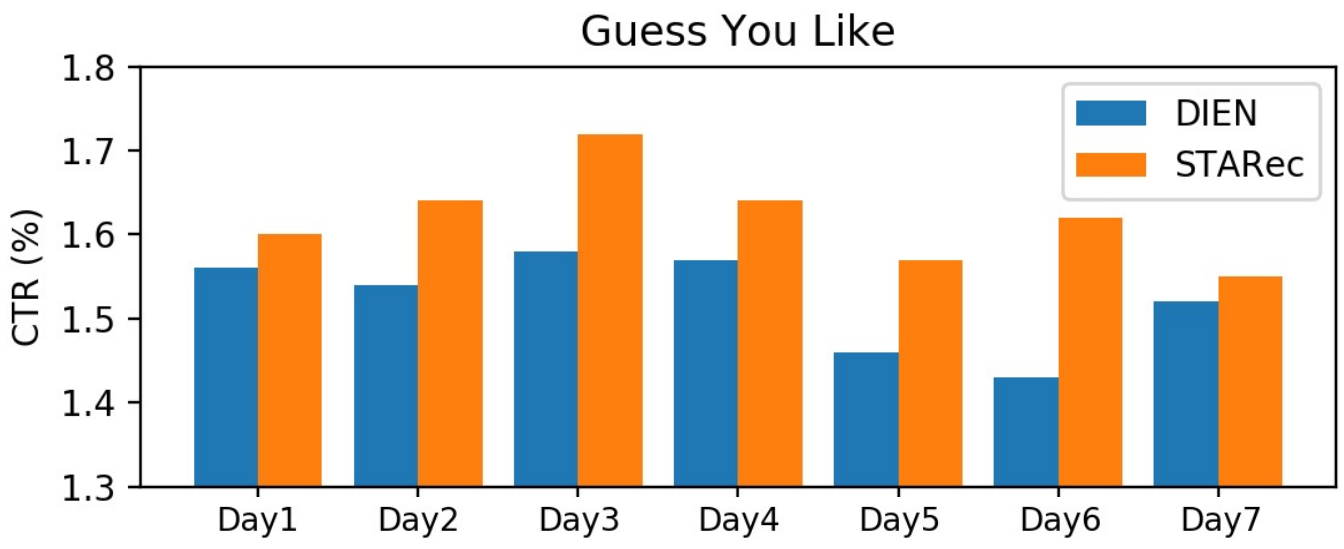}
	\vspace{-3mm}
	\caption {Daily results of online A/B test in ``Guess You Like'' scenario on CTR metric.}
	\label{fig:online}
	\vspace{-4mm}
\end{figure}

For the online experiment, we conduct A/B testing in two recommendation scenarios in Company X's online App, comparing the proposed model STARec with the current production baseline methods DIN and DIEN.
The whole online experiment lasts a week, from October 14, 2021 to October 21, 2021.
In the ``Guess You Like'' scenario, 24.1\% and 26.68\% of the users are presented with the recommendation by DIN and DIEN, respectively, while 24.2\% of the users are presented with the recommendation by STARec.
And, in ``Information Flow'', 25.4\% and 24.8\% of the users are presented with the recommendation by DIN and DIEN respectively; while 24.5\% of the users are presented with the recommendation by STARec.
We examine CTR metric defined as $\text{CTR}=\frac{\# \text{clicks}}{\# \text{impressions}}$ where \#clicks and  \#impressions are the number of clicks and impressions. 
We report the average results in Table~\ref{tab:deploy} and depict daily improvement of STARec over DIEN in Figure~\ref{fig:online} in the ``Guess You Like'' scenario.
From the table, we can see that STARec performs better in ``Guess You Like'' than ``Information Flow''.
One reason is that users' browsing lengths in ``Information Flow'' are much smaller than the lengths in ``Guess You Like'', which limits the performance of our search-based and time-aware modules.
Another reason would be that compared to ``Guess You Like'', items in ``Information Flow'' are much more diverse, which includes shopping coupons and cinema tickets, besides items in ``Guess You Like'', making searching for relevant items much harder.
From the figure, we can see the CTR improvements are rather stable where the improvement of STARec fluctuates in the range of 2\% to 13\%.

\section{Conclusion}
In this paper, we propose a novel search-based time-aware model named STARec, where we design an adaptive search-based module to retrieve relevant items and then feed this  information into a time-aware (sequential) module to capture user evolving interests.
We also design a novel label trick that allows the model to use user's previous feedbacks as the input, and reveal that this label information can significantly improve the performance.
For future work, we plan to further deploy search-based models in other real-world scenarios with sequential data. 

\minisection{Acknowledgments} 
This work is supported by China Merchants Bank Credit Card Center.
The Shanghai Jiao Tong University Team is supported by Shanghai Municipal Science and Technology Major Project (2021SHZDZX0102) and National Natural Science Foundation of China (62076161, 62177033).
We would also like to thank Wu Wen Jun Honorary Doctoral Scholarship from AI Institute, Shanghai Jiao Tong University.

\clearpage
\bibliographystyle{ACM-Reference-Format}
\bibliography{starec}
\clearpage
\appendix
\section{Pseudocode of STARec Training Procedure}
\label{app:algorithm}
In this section, we provide a detailed pseudo code of the training process in Algorithm~\ref{algo:framework}.
We analyze its time complexity as follows.
There are two main components in STARec, namely search-based and time-aware modules.
For each user-item pair (e.g., $(u_m, i_n)$), similar as analysis in \citep{qin2020user}, we need to retrieve $\widehat{T}$ (i.e., $|\widehat{\mathcal{H}}_m|$) items from the whole $T$ (i.e., $|\mathcal{H}_m|$) browsed items in item search-based module, which costs $O(1)$ time; while in user search-based module, similarly, time complexity for retrieving all relevant users (i.e., $|\widehat{\mathcal{S}}_m|$) is $O(1)$.
Hence, the overall cost for search-based module is $O(1+|\widehat{\mathcal{S}}_m|)$ for each user.
Assume that the average case time performance of recurrent units is $O(C)$.
For each user, there are $1+|\widehat{\mathcal{S}}_m|$ sequences to be fed into the time-aware sequence module.
Let $\widehat{T}$ be the average length of these sequences.
Then, the overall complexity of time-aware module is $O((1+|\widehat{\mathcal{S}}_m|)\cdot C\widehat{T})$.
Combining all the modules together, the overall complexity of STARec is $O((1+|\widehat{\mathcal{S}}_m|)\cdot (1+C\widehat{T})MN)$.

\begin{algorithm}[h]
	\caption{STARec}
	\label{algo:framework}
	\begin{algorithmic}[1]
		\REQUIRE
		dataset $\mathcal{D}=\{(u_m,i_n)\}^{M,N}_{m=1,n=1}$ with histories $\{\mathcal{H}_m\}^M_{m=1}$;
		\ENSURE
		STARec recommender with parameter $\theta$
		\vspace{1mm}
		\STATE Initialize all parameters.
		\REPEAT
		\STATE Randomly sample a batch $\mathcal{B}$ from $\mathcal{D}$
		\FOR {each data instance $(u_m, i_n)$ in $\mathcal{B}$}
		\STATE Calculate embedding vectors $\bm{e}^u_m$, $\bm{e}^i_n$ using Eq.~(\ref{eqn:fm}).
		\STATE Construct a set of relevant items $\widehat{\mathcal{H}}_m$ using Eqs.~(\ref{eqn:harditem})(\ref{eqn:searchitem}).
		\STATE Construct a set of relevant users $\widehat{\mathcal{S}}_m$ using Eqs.~(\ref{eqn:harduser})(\ref{eqn:searchuser}).
		\STATE Compute the hidden states of each sequence for $\widehat{\mathcal{H}}_m$ and $\{\widehat{\mathcal{H}}_{m'}|u_{m'}\in\widehat{\mathcal{S}}_m\}$ using Eqs.~(\ref{eqn:lstm})(\ref{eqn:attention}).
		\STATE Encode each sequence and obtain $\bm{e}_m$ and $\{\bm{e}^r_{m'}|u_{m'}\in\widehat{\mathcal{S}}_m\}$ using Eq.~(\ref{eqn:aggregation}).
		\STATE Fuse all information to generate $\widehat{y}_{mn}$ using Eq.~(\ref{eqn:prediction}).
		\ENDFOR
		\STATE Update $\theta$ by minimizing $\mathcal{L}_\theta$ according to Eq.~(\ref{eqn:loss}).
		\UNTIL convergence
	\end{algorithmic}
\end{algorithm}

\section{Experimental Configuration}
\subsection{Dataset Description}
\label{app:dataset}
We use three large-scale real-world datasets for the evaluations, and provide the detailed decription for each dataset as follows.
\begin{itemize}[topsep = 3pt,leftmargin =5pt]
\item \textbf{Tmall}\footnote{\url{https://tianchi.aliyun.com/dataset/dataDetail?dataId=42}} is a dataset consisting of 54,925,331 interactions of 424,170 users and 1,090,390 items.
These sequential histories are collected by Tmall e-commerce platform from May 2015 to November 2015 with the average sequence length of 129 and 9 feature fields. 
\item \textbf{Taobao}\footnote{\url{https://tianchi.aliyun.com/dataset/dataDetail?dataId=649}} is a dataset containing 100,150,807 interactions of 987,994 users and 4,162,024 items.
These user behaviors including several behavior types (e.g., click, purchase, add to chart, item favoring) are collected from November 2007 to December 2007 with average sequence length of 101 and 4 feature fields.
\item \textbf{Alipay}\footnote{\url{https://tianchi.aliyun.com/dataset/dataDetail?dataId=53}} is a dataset collected by Alipay, an online payment application from July 2015 to November 2015.
There are 35,179,371 interactions of 498,308 users and 2,200,191 items with average sequence length of 70 and 6 feature fields.

\end{itemize}

\subsection{Baseline Description}
\label{app:baseline}
In our paper, we compare our method against 13 strong baselines.
As STARec is proposed in the context of sequential data, most of these methods are sequential models.
We provide brief descriptions as follows.
\begin{itemize}[topsep = 3pt,leftmargin =5pt]
\item \textbf{LSTM} \citep{hochreiter1997long} is a standard long short memory approach widely used for modeling user's sequential pattern.
\item \textbf{RRN} \citep{wu2017recurrent} is a representative approach using RNN to capture the dynamic representation of users and items.
\item \textbf{STAMP} \citep{liu2018stamp} is a user action-based prediction, which models user general preference and current interest.
\item \textbf{Time-LSTM} \citep{zhu2017next} is a extension of LSTM, which considers time intervals in sequence by the time gates. 
\item \textbf{NHP} \citep{mei2016neural} is a neural Hawkes process approach which uses a self-modulating multivariate point process to model user behaviors.
\item \textbf{DUPN} \citep{ni2018perceive} is a representative learning method, which shares and learns in an end-to-end setting across user's multiple behaviors.
\item \textbf{NARM} \citep{li2017neural} is a sequential recommendation model, which uses an attention mechanism to model influence of user behaviors. 
\item \textbf{ESMM} \citep{ma2018entire} employs a feature representation transfer learning strategy over user's various behaviors.
\item \textbf{ESM$^2$} \citep{wen2019entire} designs a user behavior decomposition to model user's  various behaviors.
\item \textbf{MMoE} \citep{ma2018modeling} is a neural-based algorithm for user modeling by sharing the expert sub-models across various behaviors.
\item \textbf{DIN} \citep{zhou2018deep} designs a local activation unit to capture user interests from historical behaviors.
\item \textbf{DIEN} \citep{zhou2019deep} is an extension of DIN which captures user's evolving interests from historical behavior sequence.
\item \textbf{SIM} \citep{pi2020search} is search-based user interest model, which extracts user interests with general search units and exact search unit.
\end{itemize}
In order to further investigate the effect from each component of STARec, we design the following three variants:
\begin{itemize}[topsep = 3pt,leftmargin =5pt]
\item \textbf{STARec} is our model without using user previous feedbacks for fair comparsion.
\item \textbf{STARec}$^-_\mathtt{time}$ is a variant of STARec using a standard LSTM as the time-aware (sequential) module.
\item \textbf{STARec}$^-_\mathtt{recent}$ is a variant of STARec where $\mathcal{H}^\mathtt{RECENT}_m$ is not included in $\widehat{\mathcal{H}}_m$ (see Eq.~(\ref{eqn:harditem})).
\item \textbf{STARec}$^+_\mathtt{label}$ is a variant of STARec using user's previous feedbacks as input.
\end{itemize}

\subsection{Experimental Setting}
\label{app:config}
We split the datasets using the timestep.
For simplicity, let $T$ denote the sequence length of user browsing logs.
The training dataset contains the 1st to ($T-2$)-th user behaviors, where we use 1-st to ($T-3$)-th user records to predict the user behavior at $T-2$.
In validation set, we use 1-st to ($T-2$)-th user records to predict ($T-1$)-th user behavior, and in the test set, we use 1-st to ($T-1$)-th behaviors to predict $T$-th behavior.
The learning rate is decreased from the initial value $1\times 10^{-2}$ to $1\times 10^{-6}$ during the training process.
The batch size is set as $100$.
The weight for L2 regularization term is $4\times 10^{-5}$.
The dropout rate is set as $0.5$.
The dimension of embedding vectors is set as $64$.
All the models are trained under the same hardware settings with 16-Core AMD Ryzen 9 5950X (2.194GHZ), 62.78GB RAM, NVIDIA GeForce RTX 3080 cards.
Note that the major difference of experiment settings between our paper and \citep{qin2020user} is that we directly use click signals in the raw data as the positive feedbacks, and the negative instances are those not clicked items; while \citet{qin2020user} regards the last item as the instance receiving positive feedbacks, and randomly sample items that do not appear in the dataset as the negative samples.


\section{Deployment Discussion}
\label{app:deploy}
In this section, we introduce our hands-on experience of implementing STARec in the display advertising system with top-K recommendation and learning-to-rank tasks in Company X.
As industrial recommender or ranker systems need to process massive traffic requests per second, it's hard to make a long-term sequential user interest model serving in real-time industrial system.
As discussed in \citep{pi2019practice,pi2020search}, the storage and latency constraints could be main bottlenecks.

\subsection{Extension to Ranking Task}
In seeking to reduce the computation costs, we begin with clarifying two aforementioned tasks, namely top-K recommendation and learning-to-rank tasks.
As introduced in \emph{Definition~\ref{def:recommendation}}, the original task is a point-wise recommendation, which aims to generate similarity score for each given user-item pair.
However, in the real-world scenario, top-K recommender and ranker systems are always required to provide a list of $K$ items for each user, whose formal definition is provided as follows.

\begin{definition}
\textbf{Top-K Ranker or Recommender System}\footnote{This definition is proposed based on, and shares the same notations with \emph{Definition~\ref{def:recommendation}}.}.
Given a tuple $\langle \mathcal{U}, \mathcal{I}, \mathcal{C}, \mathcal{Q} \rangle$ where $\mathcal{Q}$ is the set of queries in ranking, the goal of the top-K ranker or recommender system is to provide a list of $K$ items $\mathcal{L}_m=\{i_1,i_2,\ldots,i_K\}$ where $i_k\in\mathcal{I}$ for each user $u_m\in\mathcal{U}$ starting at a future time $T+1$.
\end{definition}

One principle way is to first calculate the similarity for each item and then rank candidate items at descending order of their similarities.
However, the complexity of this approach prevents it to serve online, which mainly boils down to following reasons.
\begin{itemize}[topsep = 3pt,leftmargin =5pt]
\item \textbf{(R1)} As there are numerous items of various categories, our search-based module, which treats item category as key for search, needs to run multiple times, leading to high time computation.
\item \textbf{(R2)} As existing search-based model \citep{pi2020search} chooses the hard-search to save computation cost, thus, it's a great challenge to efficiently deploy our adaptive search-based module to the online system.
\end{itemize}
To mitigate this issue, we provide the following solutions.

\minisection{Mapping Queries/Demands to Certain Categories}
For the first issue, we consider to reduce the scope of candidate item categories that users may be interested in. 
In light of this, we introduce a mapping function building relation between user queries and item categories, namely a mapping model $f_\mathtt{MAP}: \mathcal{Q}\rightarrow \mathcal{C}$.
For example, in Figure~\ref{fig:motivation}, a teen $u_1$ would type ``lipstick'' in the search box, and then $f_\mathtt{MAP}$ returns category ``cosmetics''.
In this case, we only need to search and retrieve those items with cosmetics category for user $u_1$, which can significantly reduce the computation cost of searching and retrieving. 
Moreover, in some recommendation scenario lack of query information, we are also able to construct a mapping model, whose input is users' recent historical records and output is several item categories that users may be interested in, namely $f_\mathtt{MAP}:\mathcal{H}\rightarrow \mathcal{C}$.
Taking Figure~\ref{fig:motivation} as an instance, after viewing $u_1$'s recent browsed items $\mathcal{H}_1$, $f_\mathtt{MAP}$ would return category ``cosmetics'', as most of her recent interests lie in cosmetics.

\minisection{Saving Latency by Periodic Update}
As introduced in \citep{pi2020search}, one practical way is to conduct the hard-search strategy, which is a trade-off between performance gain and resource consumption.
We argue that the soft-search in our search-based module is based on similarities among embedding vectors of item categories instead of item themselves, which is much easier to learn and efficient to compute.
Besides this, we also provide a periodic update approach. 
Our approach share the same spirit with \citep{pi2020search} to build an two-level structured index for each user in an offline manner to save online latency.
Based on this structure, we also pre-compute, store, and periodically update those relevant item categories $c^i_{n'}$ satisfying $f^i_\mathtt{ADA}(c^i_{n'},c^i_{n},\bm{x}^{i1}_{n'},\bm{x}^{i1}_{n})\geq\epsilon$ for each item category $c^i_n$.
Considering that involving relevant users cause slight performance gain but huge computation cost, we choose to not include this part of STARec in our deployment. 

\subsection{Extension to Delayed Feedback}
We reveal that another issue in practice is delayed feedback caused by heavy traffic in the online system.
Formally, several labels in a user's (e.g., $u_m$'s) retrieved historical records $\widehat{\mathcal{H}}_m$ would be missing. 
Consider that this issue would be amplified, as STARec explicitly includes the user feedbacks in the proposed label tricks.
We propose to use \emph{predicted labels} generated from our model to replaced those missing \emph{original labels}.


\end{document}